\documentclass[aps,prd,onecolumn,groupedaddress,showpacs,nofootinbib,amssymb]{revtex4}
\usepackage{graphicx,bm,color}
\usepackage{amsmath}
\usepackage{amssymb}
\usepackage{amsfonts}

\newcommand{\be}{\begin{equation}}
\newcommand{\ee}{\end{equation}}
\newcommand{\bea}{\begin{eqnarray}}
\newcommand{\eea}{\end{eqnarray}}
\newcommand{\beaa}{\begin{eqnarray*}}
\newcommand{\eeaa}{\end{eqnarray*}}

\newcommand{\nn}{\nonumber \\}
\newcommand{\e}{\mathrm{e}}

\usepackage{bm}

\DeclareMathOperator{\Ei}{Ei}

\newcommand{\f}[2]{\frac{#1}{#2}}

\newcommand{\Rt}{\right}
\newcommand{\Lt}{\left}

\newcommand{\n}{\nonumber}

\allowdisplaybreaks[4] 

\begin{document}

\title{Classical Dimensional Transmutation and 
Renormalization in Massive $\lambda \phi^4$ Model}

\author{
Hiroshi Yoda$^a$ and 
Shin'ichi~Nojiri$^{a,b}$
}

\affiliation{
$^a$ Department of Physics, Nagoya University, Nagoya 464-8602, Japan \\
$^b$ Kobayashi-Maskawa Institute for the Origin of Particles and the Universe,
Nagoya University, Nagoya 464-8602, Japan
}

\begin{abstract}

Recently,  Dvali, Gomez, and Mukhanov have investigated a classical $\lambda \phi^4$ model with external source and without mass and they have clarified that there are underlying renormalization group structure, including the phenomenon of the dimensional transmutation, at purely classical level. 
Especially when the coupling $\lambda$ is negative, the classical beta function shows the property of asymptotic freedom as in QCD
\footnote{
The perturbative beta function was first written
down in \cite{Goldberger:2004jt}. 
}.   
In this paper, we investigate the $\lambda \phi^4$ model with mass, and clarify the role of the mass. 
The obtained classical beta function is identical with that of the massless $\lambda \phi^4$ model up to the corrections of the ratio of the IR cutoff to UV cutoff,
and describes the renormalization flow same as the massless theory.
We also found that the dynamically generated scale of massive theory is larger than that of massless theory, which could be due to the screening effect of the mass term.

\end{abstract}
\maketitle

\section{Introduction}

In \cite{Dvali:2011uu}, by investigating a classical $\lambda \phi^4$ model with external source, 
it has been clarified that there is a renormalization group structure, including the 
phenomenon of the dimensional transmutation, at purely classical level. 
Especially when the coupling $\lambda$ is negative, the classical beta function shows the 
property of asymptotic freedom as in QCD.  
In the paper \cite{Dvali:2011uu}, massless $\lambda \phi^4$ model has been investigated. 
In this paper, we investigate the $\lambda \phi^4$ model with mass, and clarify 
the role of the mass. 
The obtained beta function is identical with 
that of the massless $\lambda \phi^4$ model in \cite{Dvali:2011uu} up to the corrections 
of $1/\Delta$. Here $\Delta$ is the ratio of the IR cutoff to UV cutoff. 
We also, perturbatively, found that the dynamically generated scale of massive theory is larger
than that of massless theory, which could be due to the screening effect of the mass term.
Furthermore, we see the same effect and the renormalization flow in non-perturbative analyses by introducing the effective mass.

\section{Perturbative Analyses}

We start with the following Lagrangian: 
\begin{align}
\label{I}
\mathcal{L} =\frac{1}{2} \partial_\mu \phi \partial^\mu \phi 
 - \frac{1}{2} m_0^2 \phi^2 - \frac{\lambda_0}{4} \phi^4\, .
\end{align}
We also add a source term:
\begin{align}
\label{II}
J\delta^{(3)}(\bm{x}) = 4\pi Q \phi \delta^{(3)}(\bm{x}) \, ,
\end{align}
and consider spherically symmetric solution of the classical 
equation of motion with an external source of charge $Q$: 
\begin{align}
\label{classEq}
\f{1}{r^2}\f{d}{dr}\Lt (r^2\f{d\phi}{dr}\Rt ) - m_0 ^2 \phi 
 -\lambda _0 \phi ^3=-4\pi Q\delta ^{(3)}(\bm{x}) \, .
\end{align}
We expect that the mass term could screen the charge $Q$ because of opposite sign 
of the negative $\phi^4$ interaction.

We assume the solution has a form of the damping solution as in the Yukawa potential: 
\begin{align}
\phi (r)=\f{Q\tilde{f}(r,m_0) }{r}\ \e^{-m_0 r} \, .
\end{align}
Then the differential equation can be rewritten in the form of the integral equation: 
\begin{align}
\tilde{f}(r,m_0)=& 1+\alpha _0  \int ^r_{r_0} dr' \e ^{2m_0r'} 
\int ^{\infty }_{r'} \!\!\!dr'' \f{\e ^{-4m_0r''}\tilde{f}^3(r'',m_0)}{{r''}^2}
 -\tilde{N}(\alpha _0,m_0) \label{f_tilde(r)} \\ 
=&1+\alpha _0\Lt\{\int  ^r _{r_0} \!\!\! dr' \f{\e ^{-4m_0r'}(e^{2m_0 r'}
 -\e^{2m_0 r})}{2m_0{r'}^2}\tilde{f}^3(r',m_0) \Rt.\n\\
& \Lt. + \left(\e^{2m_0 r}-\e^{2m_0 r_0}\right)\int ^\infty _{r_0} dr' 
\f{\e ^{-4m_0r'}}{2m_0{r'}^2}\tilde{f}^3(r',m_0)\Rt\}
 -\tilde{N}(\alpha _0,m_0)\,.
\label{inteq}
\end{align}
Here, we have introduced an ultraviolet cutoff $r_0$ and $\alpha$ is 
defined by $\alpha _0:=-\lambda _0 Q^2 >0$.
We should note that In the limit of $m_0\to 0$, Eq.~(\ref{f_tilde(r)}) reproduces the expression 
of massless $\lambda \phi^4$ model in \cite{Dvali:2011uu} up to the constant term:
\begin{align}
\tilde{N}(\alpha _0,m_0=0)=N(\alpha _0)
 -\alpha_0 r_0\int ^\infty _{r_0} dr' \f{1}{{r'}^2}\tilde{f}^3(r',m_0=0) \, .
\end{align}
The $\mathcal {O} (\alpha _0)$ part of the solution for (\ref{f_tilde(r)}) is 
explicitly given by 
\begin{align}
\tilde{f}(r,m_0)= 1 +\alpha _0 ( 2\e^{2m_0  r}\Ei (-4m_0  r)-\Ei (-2m_0  r)  )
+ \mathcal {O} (\alpha _0 ^2) \,, 
\end{align} 
with 
\begin{align}
\tilde{N}(\alpha _0,m_0 )=\alpha _0 (2 \e^{2m_0 r_0}\Ei (-4m_0 r_0)- \Ei (-2m_0 r_0 )) 
+ \mathcal {O} (\alpha _0 ^2) \,.
\end{align}
Here the exponential integral $\Ei(x)$ is defined by
\begin{align}
\Ei(x):=-\int^\infty_{-x} dt\f{\e^{-t}}{t}\,.
\end{align}
Unfortunately, we cannot obtain analytic form of the $\mathcal {O} (\alpha _0 ^2)$ part of 
the solution by using Eq.~(\ref{inteq}) because we cannot integrate the equation analytically. 
Therefore, we now consider the expansion with respect to the mass assuming $m_0\ll 1$. 
Although the original equation (\ref{inteq}) is infrared finite, by the expansion with respect 
the mass $m_0$, there appears the infrared divergence and therefore we need to 
introduce {\it infrared cutoff} $R$ as given by
\begin{align}
\Lt. \f{d\tilde{f}}{dr}\Rt| _R=
-\alpha _0 \e ^{2m_0R} \int ^{\infty }_{R} dr'' \f{\e ^{-4m_0r''}
\tilde{f}^3(r'',m_0)}{{r''}^2}=0 \,,
\end{align}
which leads to the following expression: 
\begin{align}
\tilde{f}(r,m_0,R)=
&1+\alpha _0\Lt\{\int  ^r _{r_0} dr' \f{\e ^{-4m_0r'}(e^{2m_0 r'}
 - \e^{2m_0 r})}{2m_0{r'}^2}\tilde{f}^3(r',m_0,R) \Rt.\n\\
& \Lt. +(\e^{2m_0 r}-\e^{2m_0 r_0})\int ^R _{r_0} 
dr' \f{\e ^{-4m_0r'}}{2m_0{r'}^2}\tilde{f}^3(r',m_0,R)\Rt\} 
 -\tilde{N}(\alpha _0,m_0,R)\,.\label{f_tilde(r)ren} 
\end{align}
Furthermore we impose the renormalization condition for the IR cutoff in the 
same way as in case of the massless $\lambda\phi^4$ model in \cite{Dvali:2011uu}:
\begin{align}
\tilde{f}(R,m_0,R)=1-\tilde{N}(\alpha _0,m_0,R) \, .
\end{align}
Then Eq.~(\ref{f_tilde(r)ren}) gives
\begin{align}
\int^R_{r_0} dr' \f{\e ^{-4m_0r'}(\e^{2m_0 r'}-\e^{2m_0 r_0})}{2m_0{r'}^2}
\tilde{f}^3(r',m_0,R)=0 \,,
\end{align}
and we obtain
\begin{align}
\tilde{f}(r,m_0,R)=
&1+\alpha _0 \int^R _r dr' \f{\e ^{-4m_0r'}(\e^{2m_0 r} - \e^{2m_0 r'})}
{2m_0{r'}^2}\tilde{f}^3(r',m_0,R) -\tilde{N}(\alpha _0,m_0,R)\,.
\label{A1}
\end{align}
We now use new variable $t$ and new parameter $\Delta$ by $t :=r/r_0$ 
and $\Delta:=R/r_0$. Furthermore we choose $R=1/m_0$. 
Then the expression (\ref{A1}) reduces to 
\begin{align}
\tilde{f}(t,\Delta )=1+\Delta\alpha _0 \int^{\Delta} _t 
dt' \f{\e^{-4t'/\Delta}\left(\e^{2t/\Delta} - \e^{2t'/\Delta}\right)}
{2{t'}^2}\tilde{f}^3(t' ,\Delta ) -\tilde{N}(\alpha _0 ,\Delta)\,. 
\label{f_tilde(t)}
\end{align}
As in case of Eq.~(\ref{f_tilde(r)}), we cannot solve (\ref{f_tilde(t)}) 
analytically. 
Since $\Delta$ is the ratio of IR cutoff to UV cutoff, $\Delta$ should be large. 
So we expand the integrand in (\ref{f_tilde(r)}) by the power of $1/\Delta$, 
and neglect the $\mathcal {O} \left(\left(1/\Delta\right)^2\right)$ terms.
Therefore we find 
\begin{align}
\tilde{f}(x ,\Delta )=1&+\alpha_0 \int^x _{\ln \Delta} dx' \Lt\{ 
1- \e^{x-x'}+\f{1}{\Delta} \e^x\Lt( 4 - \e^{x-x'} - 3 \e^{-(x-x')} \Rt) \Rt\} 
\tilde{f}^3(x' ,\Delta ) -\tilde{N}(\alpha _0,\Delta)\,.
\end{align}
Now we have used a new variable $x$ defined by $x:=\ln t=\ln r/r_0$. 
We further change the variable by $x\rightarrow y:=x-\ln \Delta=\ln (m_0 r)$ 
and find the following expression: 
\begin{align}
\tilde{f}(y) = 1&+\alpha _0 \int^y_0 dy' \Lt\{ 1 - \e^{y-y'} 
+ \e^y\Lt( 4 - \e^{y-y'} - 3 \e^{-(y-y')} \Rt) \Rt\} \tilde{f}^3(y')
 -\tilde{N}(\alpha _0 ) \, .\label{f_tilde(y)}
\end{align}
The recursive solution of (\ref{f_tilde(y)}) is given by
\begin{align}
\tilde{f}(y) = 1+\alpha_0 \left( y - (3-4 y) \e^y - \e^{2 y}\right) 
+\alpha_0^2 \left( 3y + \frac{3}{2} y^2 + 15 (5+y) \e^y 
+ \left( 66-39 y + 6 y^2\right) \e^{2 y}\right) 
+\mathcal {O} (\alpha_0 ^3) \, ,\label{f_tilde(y)rec}
\end{align}
namely,
\begin{align}
\tilde{f}(r) =& 1 + \alpha _0\left[\ln (m_0 r) -m_0 r (3-4 \ln (m_0 r))-(m_0 r)^2 \right] \nn
& +\alpha _0^2 \left[ 3\ln (m_0 r)+\frac{3}{2} \ln^2(m_0 r) + 15 m_0 r (5 + \ln (m_0 r)) 
+ (m_0 r)^2 \left( 66-39 \ln (m_0 r)+6 \ln ^2(m_0 r)\right)\right] 
+\mathcal {O} (\alpha_0 ^3) \, ,
\end{align}
with
\begin{align}
\tilde{N}(\alpha _0)=4 \alpha_0 -9 \alpha_0 ^2+\mathcal {O} (\alpha_0 ^3)\,.\label{N_tilde_rec}
\end{align}
In (\ref{f_tilde(y)rec}), we have neglected $\e^y$ terms, which lead to identify $y$ with $x$. 
The obtained equation is identical with that in case of massless $\lambda \phi^4$ model 
in \cite{Dvali:2011uu} although the expression of (\ref{N_tilde_rec}) is different.   

We now define an effective coupling constant $\alpha_\mathrm{eff}(y)$ by 
$\alpha_\mathrm{eff}(y):=\alpha_0\tilde{f}^2(y)$. 
Then by using (\ref{f_tilde(y)rec}), we obtain
\begin{align}
\alpha_\mathrm{eff}(y)=&\alpha_0 \Lt \{ 1+2 \alpha \left[(y - \e^y (3-4 y) 
 - \e^{2 y}\right] \Rt. \n\\ 
& +\alpha_0 ^2\left[ 6 y +4 y^2 +2 \left(75+12 y+4 y^2\right) \e^y 
\Lt. -\left(123-52y-4 y^2\right) \e^{2 y} + (6-8 y) \e^{3 y} 
+ \e^{4 y}\right ]+\mathcal {O} (\alpha_0 ^3)  \Rt \} \\
=:&\sum^{\infty}_{n=0}\alpha_0 ^{n+1}\tilde{g}_n(y) \,,
\end{align}
thus,
\begin{align}
\alpha_\mathrm{eff}(r)=&\alpha_0 \Lt \{ 1
+2 \alpha \left[(\ln (m_0 r)-m_0 r (3-4 \ln (m_0 r))-(m_0 r)^2 \right] \Rt. \n\\ 
& +\alpha_0 ^2\left[ 6 \ln (m_0 r) +4 \ln ^2 (m_0 r) 
+2m_0 r \left(75+12 \ln (m_0 r)+4 \ln ^2(m_0 r)\right)  \Rt. \n\\
& \Lt.\Lt. -(m_0 r)^2 \left(123-52\ln (m_0 r)-4 \ln ^2(m_0 r)\right) 
+ (m_0 r)^3 (6-8 \ln (m_0 r))+(m_0 r)^4 \right ]+\mathcal {O} (\alpha_0 ^3)  \Rt \} \,.
\end{align}

If we choose $\alpha_0=\alpha_\mathrm{eff}(r_0)$ and fix $\Delta$,that is the ratio of IR cutoff 
to UV cutoff, $\tilde{f}(y)$ only depends on $r_0$ logarithmically.
So, we are able to define {\it classical} beta function as in a way similar to 
\cite{Dvali:2011uu}, 
\begin{align}
\beta :=\f{\alpha_\mathrm{eff}}{dx}=\f{\alpha_\mathrm{eff}}{dy}
=\alpha_\mathrm{eff}^2\sum^{\infty}_{k=0}\tilde{g}'_k(-\ln \Delta )\alpha_\mathrm{eff} ^{k}\, .
\label{A2}
\end{align}
In (\ref{A2}), we have put $y=-\ln \Delta$, that is, $x=y + \ln \Delta = 0$.
By neglecting $\mathcal {O} \left( \left(1/\Delta\right)^2 \right)$ terms, we obtain
\begin{align}
\beta (\alpha )=& \left[2+ \f{2}{\Delta} + \f{8}{\Delta} \ln\Lt(\f{1}{\Delta}\Rt)\right] \alpha ^2 
+\left[ 6 + \f{174}{\Delta} + 8\left( 1+ \f{5}{\Delta}
+\f{1}{\Delta} \ln\Lt(\f{1}{\Delta}\Rt)\right) \ln\Lt(\f{1}{\Delta}\Rt)\right ]\alpha ^3
+\mathcal {O} (\alpha ^4) \, . \label{beta}
\end{align}
We should note, now, $\alpha :=\alpha _\mathrm{eff}$.

In (\ref{beta}), numerical coefficients of each powers of $\alpha$ are equal to the case of the massless theory
\footnote{We have checked the terms to $\mathcal {O} (\alpha ^8)$.}, 
but we cannot take the massless limit, $\Delta\rightarrow \infty$, because of logarithmic divergence of $\mathcal {O} (\alpha ^3)$ terms.

We check the {\it 1-loop} part, that is, $\mathcal{O} \left(\alpha^2\right)$ part 
of the classical beta function (\ref{beta}),
\begin{align}
\beta (\alpha )=\f{d\alpha}{dx}=2\kappa \alpha ^2  \, . \label{beta1}
\end{align}
Here
\begin{align}
\kappa (1/\Delta ):=1+\f{1}{\Delta} + \f{4}{\Delta} \ln\Lt(\f{1}{\Delta}\Rt) \,  ,
\end{align}
and we should note 
\begin{align}
0<\kappa <1
\end{align}
for $\Delta >10$. 
We now assume $\Delta$ does not depend on $r_0$ and solve (\ref{beta1}).
Then, we obtain {\it dynamically generated scale} $\tilde{R}_c$ in the same way similar to 
the case of the massless $\lambda \phi^4$ model in \cite{Dvali:2011uu}, 
\begin{align}
\tilde{R}_c=r_0 \e^{-1/2\kappa \lambda_0 Q^2}>R_c \, . 
\end{align}
$R_c$ is the dynamically generated scale in massless $\lambda \phi^4$ model \cite{Dvali:2011uu}.
Namely, the dynamically generated scale of massive theory is larger than that of massless theory. 
This result could be consistent with screening effect of the mass term.

Obviously, the physical running coupling constant $\lambda (r)$ is given by
\begin{align}
\lambda (r)=\f{1}{2Q^2\ln (\tilde{R}_c/r)} \, .
\end{align} 

We found that the dynamically generated scale of massive theory is larger than that 
of massless theory, which could be due to the screening effect of the mass term.

\section{Non-Perturbative Analyses}

We find the {\it classical} beta function (\ref{A2}) as in a way similar to \cite{Dvali:2011uu}.
So, we can obtain the differential equation of the classical beta function from the equation of motion 
(\ref{classEq}) and (\ref{A2}), as follows:
\begin{align}
\Lt( 1+\f{2 \e ^x}{\Delta}\Rt)\beta (\alpha )
=2 \exp\Lt[-\f{2 \e ^x}{\Delta}\Rt] \alpha ^2 +
\f{1}{2}\Lt( \f{d\beta (\alpha ) ^2}{d \alpha }-\f{\beta (\alpha ) ^2}{\alpha }\Rt) \label{eq_beta}\, .
\end{align} 
We should note that (\ref{eq_beta}) coincides with the equation of the massless theory 
in the massless limit.
We now consider to solve Eq.~(\ref{eq_beta}) in a non-perturbative way. 

If we choose that $x$ could be independent of $\alpha$, 
the solution of (\ref{eq_beta}) is given by
\begin{align}
\beta (\bar{\alpha} )=\Lt( 1+\f{2 \e ^x}{\Delta}\Rt) ^3 \exp\Lt[\f{2 \e ^x}{\Delta}\Rt]
\Lt( 2\bar{\alpha}^2+6\bar{\alpha}^3+48\bar{\alpha}^4+570\bar{\alpha}^5+  \cdots \Rt)\, .
\label{beta_alpha_bar}
\end{align}
Here, $\bar{\alpha}$ is defined by
\begin{align}
\bar{\alpha}:=\f{\exp\Lt[-\f{2 \e ^x}{\Delta}\Rt]}{\Lt( 1+\f{2 \e ^x}{\Delta}\Rt) ^2}\alpha 
=\f{\e^{-2 m_0 r}}{\Lt( 1+2 m_0 r\Rt) ^2}\alpha \, .
\end{align}
In (\ref{beta_alpha_bar}), the numerical coefficients of each powers of $\bar{\alpha}$ are 
equal to those in case of the massless model. 
This situation is not changed from (\ref{beta}) but the $\Delta$ dependence in the beta function in 
(\ref{beta_alpha_bar}) is different from that in (\ref{beta}).
Since $\bar{\alpha}<\alpha$, this could be interpreted as the screening effect due to the mass term.

In order to investigate the renormalization flow of the massive theory we now remove the explicit 
scale dependence from the expression in (\ref{eq_beta}). 
For this purpose, we introduce the new variable $w(x)$ corresponding to scale and 
rewrote (\ref{eq_beta}) as follows, 
\begin{align}
\Lt( 1+\f{2 \e ^x}{\Delta}\Rt) \tilde{\beta} 
=\f{2\exp\Lt[-\f{2 \e ^x}{\Delta}\Rt]}{w'}\alpha^2
+\f{w'}{2}\Lt(\f{d\tilde{\beta}^2}{d\alpha}-\f{\tilde{\beta}^2}{\alpha} 
+\f{1}{(w')^2}\f{d(w')^2}{d\alpha}\tilde{\beta}^2\Rt) 
\label{eq_beta_tilde}\, .
\end{align}
Here we have defined $w':=\f{dw}{dx}$ and $\tilde{\beta}:=\f{d\alpha}{dw}$. 
We now require
\begin{align}
\f{2 \e ^x}{\Delta} =\f{1}{2w'}\f{d(w')^2}{d\alpha}\tilde{\beta}=\f{dw'}{dw}
\end{align}
on (\ref{eq_beta_tilde}), that is, 
\begin{align}
w(x) = \Ei (2 \e ^x/\Delta) -\Lt( \gamma +\log(2/\Delta) \Rt) 
= \Ei (2 m_0 r) -\Lt( \gamma +\log(2 m_0 r_0) \Rt) \, . 
\label{Eiwx}
\end{align}
Here we have chosen one of integration constants so that we have $w(x)=x$ in the massless limit. 
Another constant can be arbitrary because it is identified with the constant factor 
of $\tilde{\beta}$. 
By using (\ref{Eiwx}), we find  
\begin{align}
w'(x)=\exp\Lt[\f{2 \e ^x}{\Delta}\Rt]\, ,
\end{align}
and we obtain 
\begin{align}
\tilde{\beta} =\f{2}{(w')^2} \alpha ^2
+\f{w'}{2}\Lt(\f{d\tilde{\beta}^2}{d\alpha}-\f{\tilde{\beta}^2}{\alpha} \Rt) \, .
\end{align}
We now define the screened coupling constant as 
\begin{align}
\tilde{\alpha}:=\f{\alpha}{w'}=\exp\Lt[-\f{2 \e ^x}{\Delta}\Rt]\alpha \, .
\end{align}
Then we obtain
\begin{align}
\tilde{\beta}(\tilde{\alpha})=2\tilde{\alpha}^2
+\f{1}{2}\Lt( \f{d\tilde{\beta}(\tilde{\alpha})^2}{d\tilde{\alpha}} 
 -\f{\tilde{\beta}(\tilde{\alpha})^2}{\tilde{\alpha}} \Rt) 
 -\f{dw'}{dw}\f{d\tilde{\beta}(\tilde{\alpha})}{d\tilde{\alpha}} \tilde{\alpha} \, .
\label{eq_beta_tilde2}
\end{align}
Eq.~(\ref{eq_beta_tilde2}) is identical with the equation corresponding to the case 
of massless model up to the last term which depends on $w$ explicitly. 
For remove this $w$ dependence, we introduce the {\it effective mass} 
as $p^\mu p_\mu\phi=m_{\mathrm{eff}}^2\phi$, namely,
\begin{align}
m_{\mathrm{eff}}^2=m_0 ^2+\lambda_0 \phi ^2=m_0 ^2 
 - \f{\alpha\e^{-2m_0 r}}{r^2}=m_0 ^2\Lt\{1-4\tilde{\alpha}\Lt(\f{dw'}{dw}\Rt)^{-2}\Rt\}\, .
\end{align}
Then by using (\ref{eq_beta_tilde2}), we obtain
\begin{align}
\tilde{\beta}(\tilde{\alpha})=2\tilde{\alpha}^2
+\f{1}{2}\Lt( \f{d\tilde{\beta}(\tilde{\alpha})^2}{d\tilde{\alpha}} 
 -\f{\tilde{\beta}(\tilde{\alpha})^2}{\tilde{\alpha}} \Rt) 
 -2\sqrt{\f{m_0 ^2}{m_0 ^2-m_{\mathrm{eff}}^2}}
\f{d\tilde{\beta}(\tilde{\alpha})}{d\tilde{\alpha}}\tilde{\alpha}^{\f{3}{2}}\, .
\label{eq_beta_tilde3}
\end{align}
We regard $m_\mathrm{eff}$ as the screened physical mass by taking account for the 
$\lambda\phi ^4$ interaction.
Therefore when we like to investigate the renormalization flow of the coupling constant 
$\alpha$, we need to solve (\ref{eq_beta_tilde3}) on fixed $m_\mathrm{eff}^2$.

Although the expression of (\ref{eq_beta_tilde3}) is completely different from that 
in case of the massless model, surprisingly, for fixed $m_\mathrm{eff}^2$, 
the numerical solutions of (\ref{eq_beta_tilde3}) have the behavior identical with that in the massless model 
even if we choose the value of $\sqrt{m_0 ^2/\left(m_0^2-m_{\mathrm{eff}}^2\right)}$ large or small. 
Furthermore, in order that $\sqrt{m_0 ^2/\left(m_0^2-m_{\mathrm{eff}}^2\right)}$ in Eq.~(\ref{eq_beta_tilde3}) 
could be real, the value of $m_\mathrm{eff}^2$ is restricted to be $-\infty<m_\mathrm{eff}^2<m_0^2$ 
but there should exist the limit of $m_\mathrm{eff}^2\to +\infty$ and the solutions would be realized 
for arbitrary $m_\mathrm{eff}^2$.
For these reasons, we can take the limit $m_0\rightarrow \infty $, namely, 
$\sqrt{m_0 ^2/\left(m_0 ^2-m_{\mathrm{eff}}^2\right)}\rightarrow 1$,
and we can check that the behavior is also identical with that in the massless model in this limit.

\section{Summary}

In this paper we generalized the approach of Ref.~\cite{Dvali:2011uu} for massive theory and have investigated 
the structure of the classical renormalization group of the $\lambda \phi^4$ model with mass, 
as in case of massless $\lambda \phi^4$ model in \cite{Dvali:2011uu}. 
We find the classical beta function perturbatively as in (\ref{beta}) by solving 
the recursive equation (\ref{f_tilde(r)}) by choosing the renormalization condition as 
in \cite{Dvali:2011uu} and identifying the infrared cutoff $R$ with the inverse of the bare mass $m_0$, $R=1/m_0$.
The obtained beta function is identical with that of the massless $\lambda \phi^4$ model in \cite{Dvali:2011uu} 
up to the corrections of $1/\Delta$ which is the ratio of the IR cutoff to UV cutoff 
and these corrections can be interpreted as the screening effects due to the mass term.
Since the numerical coefficients of each powers of $\alpha$ are equal to the case of the massless theory, 
we expect that the perturbative classical beta function (\ref{beta}) could describe the correct behavior 
of the classical renormalization flow.
However, because of the divergence in the massless limit $\Delta\rightarrow \infty $, 
this perturbative beta function would be unreliable.

Therefore, we have analyzed the classical beta function non-perturbatively 
by introducing the effective mass and using Eq.~(\ref{eq_beta_tilde3}).
This equation has a structure similar to that of the equation in the massless theory, and coincides with the 
equation in the massless theory in the massless limit.
We should note, however, that Eq.~(\ref{eq_beta_tilde3}) is not the equation for $\beta(\alpha)$ 
but that for the scaled beta function and the screened coupling constant $\tilde{\beta}(\tilde{\alpha})$.
Since these scalings are monotonic, however, the solution of this equation describes the renormalization 
flow of original variables.
So, we solved Eq.~(\ref{eq_beta_tilde3}) on a fixed effective mass, and we obtained the behavior identical 
with that in the massless model numerically.

Consequently, we conclude that the classical analysis of the structure in the renormalization group 
for $\lambda \phi^4$ model such as that in Ref.~\cite{Dvali:2011uu} is even available in the massive theory,
and verify the screening effects of the mass term.

Finally, we comment on two subjects.
First, we discussed the overall structure of the renormalization group of $\lambda \phi^4$ model 
for both of negative and positive coupling constant.
We obtained the equation of classical beta function in positive coupling theory, as in the model 
of Ref.~\cite{Dvali:2011uu}, by the replacement $-\alpha$ with $\alpha$ in Eq.~(\ref{eq_beta_tilde3}).
However, this equation would not describe the structure of the renormalization group of the positive coupling theory,
because if we consider the limit of $m_0^2\rightarrow -\infty$, as in the case of the negative theory,
we should replace $-\alpha$ with $\alpha$ again, since the range of the values of $m_\mathrm{eff}^2$ 
is restricted to be $m_0^2<m_\mathrm{eff}^2<\infty$.
Nevertheless, we can obtain the equation of the positive coupling theory only when $m_\mathrm{eff}^2$ 
goes to infinity simultaneously with $m_0^2\rightarrow -\infty$,
because in this situation $m_{\mathrm{eff}}^2=m_0 ^2$, 
namely $\sqrt{m_0 ^2/(m_0 ^2-m_{\mathrm{eff}}^2)}\rightarrow \infty$, 
and we can regard the equation as $d\tilde{\beta}/d\tilde{\alpha}=0$, namely $\beta =0$.
Therefor, it would be suggested that the positive $\lambda \phi^4$ theory is trivial.
Second, it would be interesting now to consider further generalization of the model 
to curved space-time where the effective mass term $\xi R \phi^2$ appears. 
This may be considered in analogy with renormalization group approach in curved  
space (for review, see \cite{Odintsov:1990mt}). 

\section*{Acknowledgments.}

The authors are grateful to S.~D.~Odintsov for discussions. 
S.N. is also indebted to G.~Dvali and S.~Mukhanov for drawing his attention 
to the subject discussed in this paper. 
S.N. and H.Y. are supported in part by Global COE Program of Nagoya University (G07)
provided by the Ministry of Education, Culture, Sports, Science \&
Technology and S.N. is also supported by the JSPS Grant-in-Aid for Scientific Research (S) \# 22224003
and (C) \# 23540296.


\begin{thebibliography}{99}

\bibitem{Dvali:2011uu}
G.~Dvali, C.~Gomez and S.~Mukhanov,
JHEP {\bf 1112} (2011) 103  
[arXiv:1107.0870 [hep-th]].  

\bibitem{Goldberger:2004jt}
W.~D.~Goldberger and I.~Z.~Rothstein,
Phys.\ Rev.\ D {\bf 73} (2006) 104029  
[hep-th/0409156].  

\bibitem{Odintsov:1990mt}
S.~D.~Odintsov,
Fortsch.\ Phys.\  {\bf 39} (1991) 621.  


\end{thebibliography}
\end{document}